\documentclass[twoside]{dis08}
\usepackage[latin1]{inputenc}
\usepackage[dvips]{graphicx,epsfig,color}
\usepackage{wrapfig,rotating}
\usepackage{amssymb,amsmath,array}

\begin{document}
\title{Including heavy quark production in ZEUS-PDF fits}

\author{A M Cooper-Sarkar
%
%
\vspace{.3cm}\\
%
Oxford University - Dept of Physics \\
Denys Wilkinson Bdg, Oxford, OX1 3RH - UK
}

\maketitle

\begin{abstract}
At HERA heavy quarks may contribute up to $30\%$ of the structure function 
$F_2$. The potential of including heavy-quark data in the ZEUS PDF fits 
is explored, using $D^*$ double differential cross-sections as well as
 the inclusive quantities $F_2^{c\bar{c}}$, $F_2^{c\bar{c}}$. 
The introduction of heavy quarks requires an extension of the DGLAP 
formalism. The effect of using different heavy flavour number schemes, and 
different approaches to the running of $\alpha_s$, are compared.
\end{abstract}


Parton Density Function (PDF) determinations are usually global 
fits~\cite{mrst,mrst06,cteq,zeus-s}, which use inclusive cross-section data and 
structure function measurements from deep inelastic lepton hadron scattering 
(DIS) data as well as some other exclusive cross-ections. 
The kinematics
of lepton hadron scattering is described in terms of the variables $Q^2$, the
invariant mass of the exchanged vector boson, Bjorken $x$, the fraction
of the momentum of the incoming nucleon taken by the struck quark (in the 
quark-parton model), and $y$ which measures the energy transfer between the
lepton and hadron systems.
The differential cross-section for the neutral current (NC) 
process is given in terms of the structure functions by
\[
\frac {d^2\sigma(e^{\pm}p) } {dxdQ^2} =  \frac {2\pi\alpha^2} {Q^4 x}
\left[Y_+\,F_2(x,Q^2) - y^2 \,F_L(x,Q^2)
\mp Y_-\, xF_3(x,Q^2) \right],
\]
where $\displaystyle Y_\pm=1\pm(1-y)^2$. In the HERA kinematic range there 
is a sizeable 
contribution to the $F_2$ structure function from heavy quarks, 
particularly charm. Thus heavy quarks must be properly treated in the fomalism.
 Furthermore fitting data on charm production may help to give constraints 
on the gluon PDF at low-$x$.

The most frequent approaches to the inclusion of heavy quarks
 within the conventional framework of QCD evolution 
using the DGLAP equations are~\footnote{Charm production is described 
here but a similar formalism describes beauty production}:
\begin{itemize}
\item ZM-VFN (zero-mass variable flavour number schemes) in which the
charm parton density $c(x,Q^2)$ satisfies
$c(x,Q^2)=0$ for $Q^2\leq \mu^2_c$ and $n_f=3+\theta(Q^2-\mu^2_c)$
in the splitting functions and $\beta$ function. The threshold $\mu^2_c$,
which is in the range $m^2_c<\mu^2_c<4m^2_c$, is chosen
so that $F_2^c(x,Q^2)=2e_c^2xc(x,Q^2)$ gives a satisfactory description
of the data. The advantage of this approach is that the simplicity
of the massless DGLAP equations is retained. The disadvantage is that the
physical threshold $\hat{W}^2=Q^2({1\over z}-1)\geq 4m_c^2$ is not treated
correctly ($\hat{W}$ is the $\gamma^*g$ CM energy).
\item FFN (fixed flavour number schemes) in which there is no charm parton
density and all charmed quarks are generated by the BGF process. The 
advantage of the FFNS scheme is that the threshold region is correctly
handled, but the disadvantge is that large $\ln(Q^2/m^2_c)$ terms appear
and charm has to be treated ab initio in each hard process.
\item GM-VFN (general mass variable flavour number schemes), which aim to 
treat the 
threshold correctly and absorb $\ln(Q^2/m^2_c)$ terms into a charm parton
density at large $Q^2$. There are differing versions of such 
schemes~\cite{cteq65,hqnew}
\end{itemize}

For the ZEUS PDF analyses~\cite{zeusj,zeus-s}, 
the heavy quark production scheme used was the
general mass variable flavour number scheme of Roberts and Thorne 
(TR-VFN)~\cite{hq,hqnew}. However we also investigated the use of the FFN 
for 3-flavours with the renormalisation and factorisation scale 
for light quarks both set to $Q^2$ but the factorisation scale for heavy 
quarks set to $Q^2 + 4 m_c^2$. The reason for these choices of scheme and scale
 is that these are the choices made in the programme 
HVQDIS~\cite{hvqdis3} which was used 
to extract $F_2^{c\bar{c}}$ from data on $D^*$ production. Furthermore, it has recently become evident 
that the use of the FFN scheme implies a treatment of the running of $\alpha_s$
which is different from that of the VFN schemes. 
In VFN schemes $\alpha_s$ is matched at flavour 
thresholds~\cite{match}, but the slope of $\alpha_s$ is discontinuous. 
In the FFN scheme we must use 
a 3-flavour $\alpha_S$ which is continuous in $Q^2$. This requires 
an equivalent value of $\alpha_s(M_Z)=0.105$ in order to be consistent, 
at low $Q^2$, with the results of using a value of $\alpha_s(M_Z)=0.118$ in 
the usual VFN schemes.

\begin{figure}[tbp]
\vspace{-0.5cm} 
\centerline{
\epsfig{figure=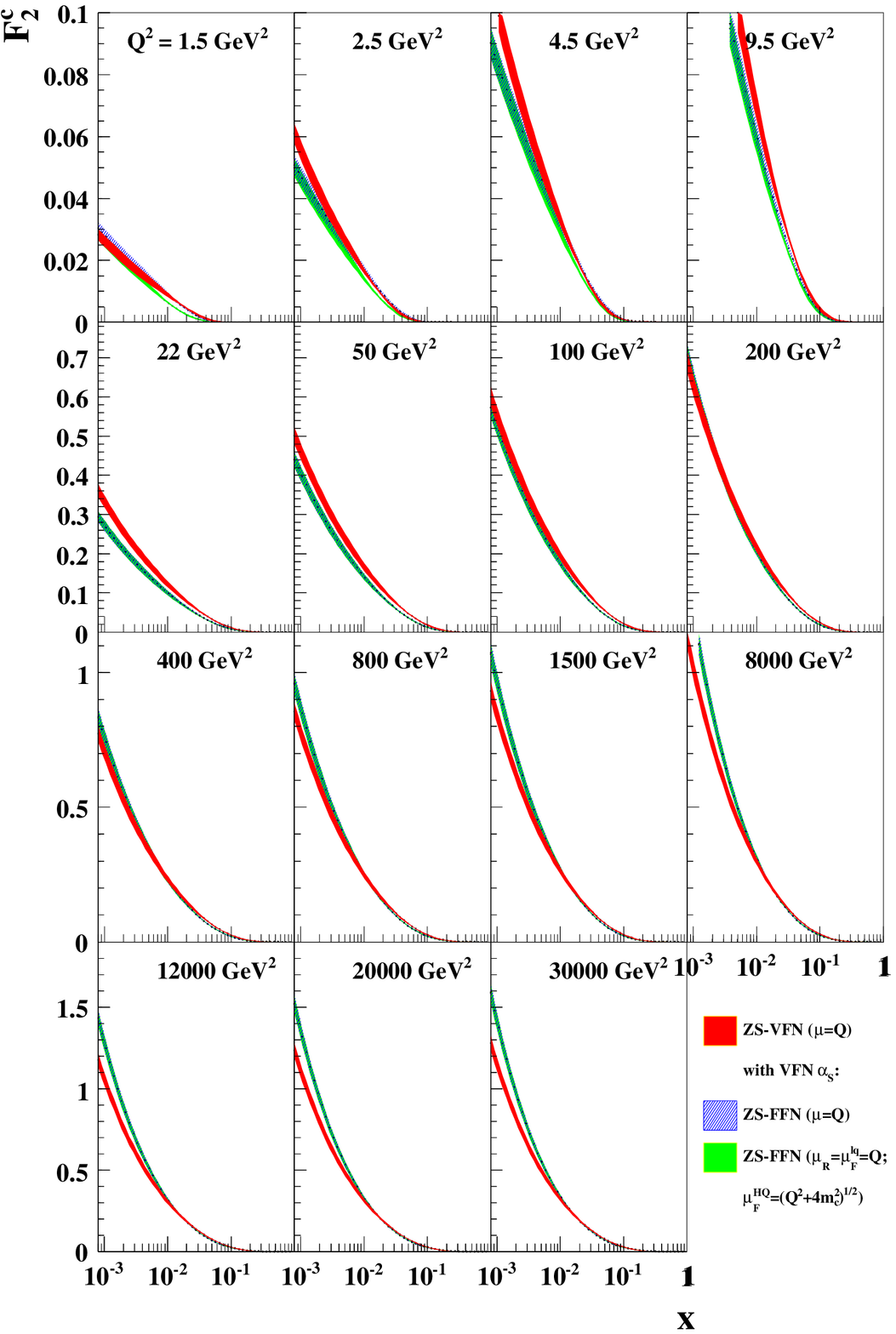,width=0.5\textwidth,height=5.5cm}
\epsfig{figure=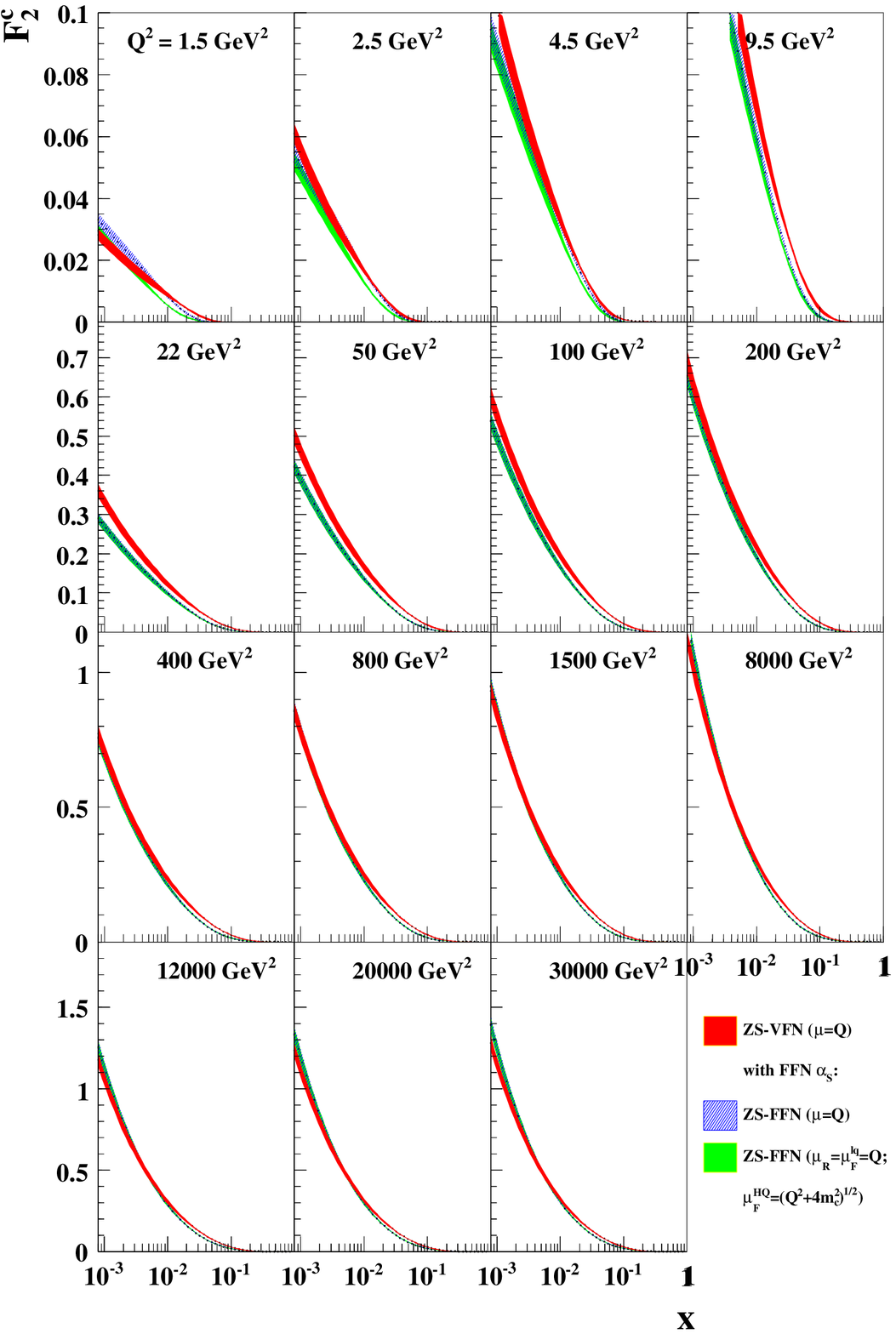,width=0.5\textwidth,height=5.5cm}}
\caption { Comparison of predictions for $F_2^{c\bar{c}}$, from fits which 
use the GM-VFN scheme and the FFN scheme with two different factorisation 
scales: on the left hand side the FFN schemes still use a VFN treatment of
$\alpha_s$, whereas on the right hand side a 3-flavour $\alpha_s$ is used.}
\label{fig:f2calphas}
\end{figure}

In Fig~\ref{fig:f2calphas} we compare different heavy quark factorisation 
scales and different treatments of the running of 
$\alpha_s$ for predictions of $F_2^{c\bar{c}}$. 
We see that within the FFN scheme the choice of the heavy quark factorisation 
scale makes only a small difference at low $Q^2$. 
The treatment of $\alpha_S$ gives larger differences. The FFN scheme and 
GM-VFN scheme differ for almost all $Q^2$ if $\alpha_S$ runs as for the VFN 
schemes. However if a 3-flavour $\alpha_S$ is applied in the FFN schemes there 
is much better agreement of all schemes at higher $Q^2$.

\begin{figure}[tbp]
\vspace{-0.5cm} 
\centerline{
\epsfig{figure=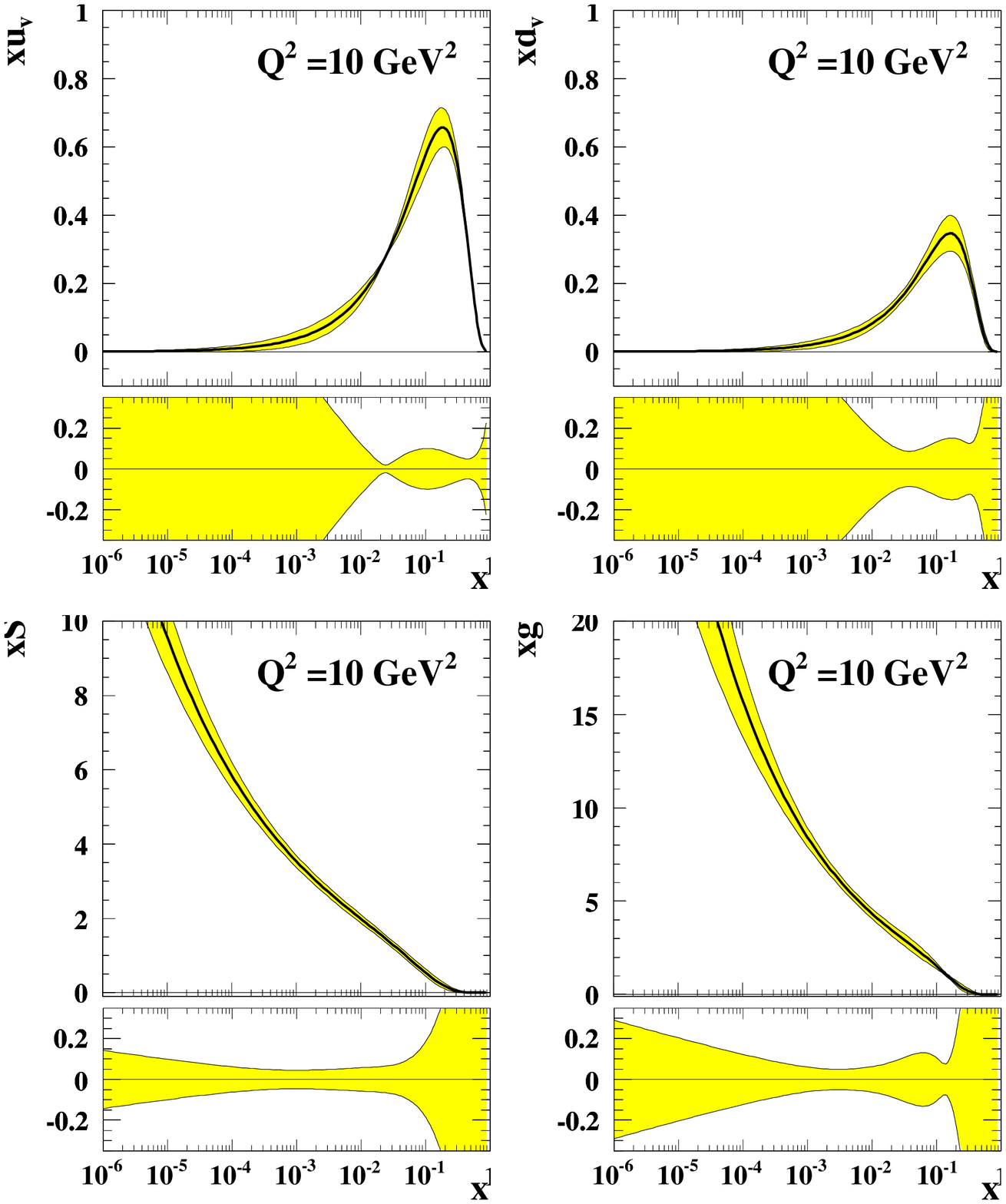,width=0.5\textwidth,height=5.0cm}
\epsfig{figure=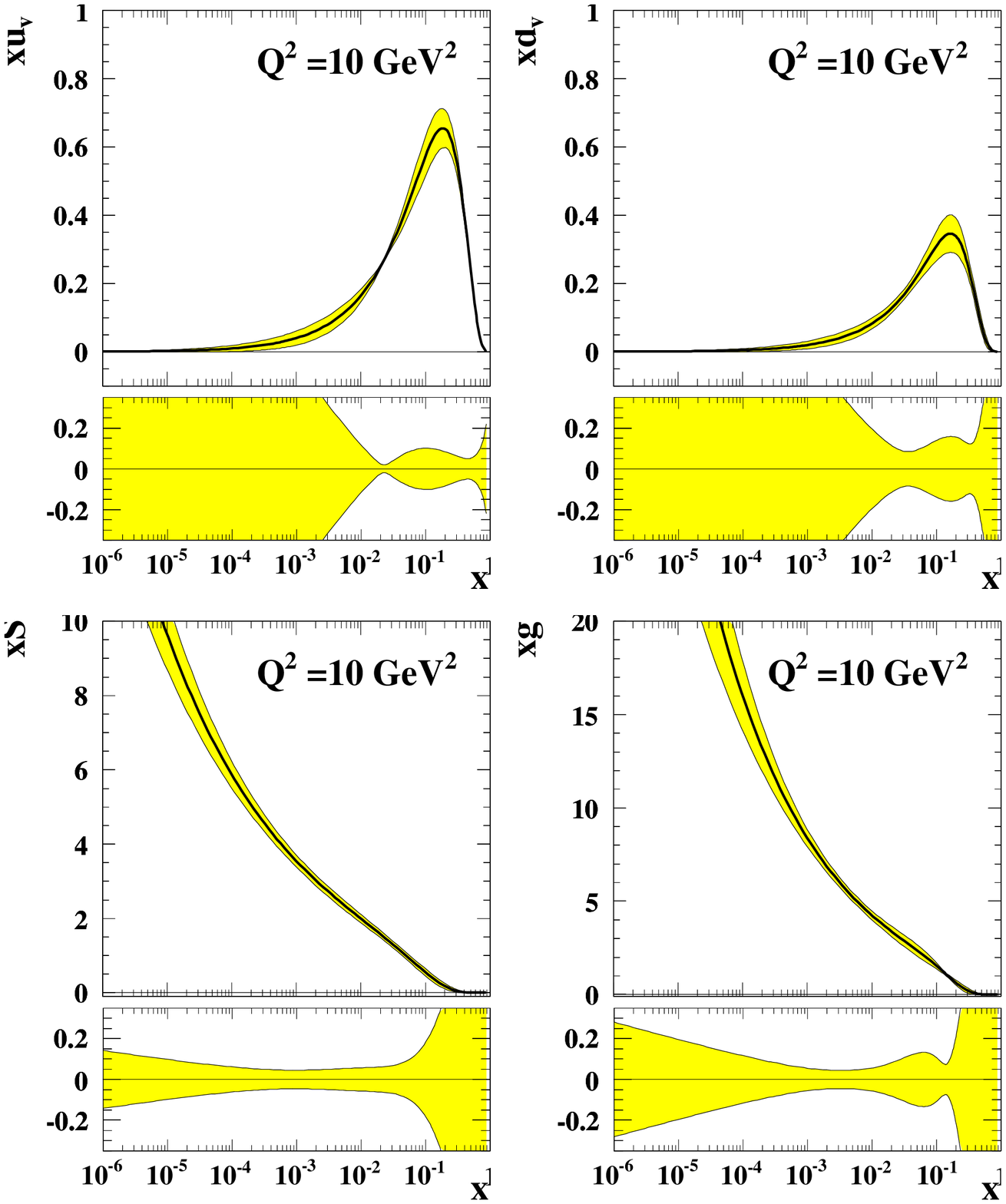,width=0.5\textwidth,height=5.0cm}}
\caption { The $u$-valence, $d$-valence, Sea and gluon PDFs and 
their fractional uncertainties at $Q^2= 10$GeV$^2$, 
from a) the ZEUS-pol PDF fit (left) and b) a smilar fit with 
$F_2^{c\bar{c}}$ and $F_2^{b\bar{b}}$ data included (right).}
\label{fig:f2charm}
\end{figure}

There is now new data on  $F_2^{c\bar{c}}$~\cite{hera2charm} and 
$F_2^{b\bar{b}}$~\cite{hera2beauty} from HERA-II running to add to older 
the HERA-I charm data~\cite{dstar2003}. To investigate the potential of 
these data to constrain the gluon PDF, we used the ZEUS-pol PDF
fit~\cite{kunihiro} and added the charm data. The new 
data do not influence the central values of the ZEUS-pol fit significantly.
However there is a small improvement in the precision of the low-$x$ gluon.
Fig.~\ref{fig:f2charm} compares the PDFs and their uncertainties, as 
extracted from the ZEUS-pol PDF fit,  
with the those extracted from a similar fit  
including the  $F_2^{c\bar{c}}$  and $F_2^{b\bar{b}}$ data.  
This illustrates that the charm data has the potential to constrain the 
gluon PDF uncertainties.
 
We have also compared fits using 
the GM-VFN formalism with those using the FFN formalism with 3-flavour 
$\alpha_s$. When using the FFN 
formalism one should not really use high-$Q^2$ data, because 
large $\ln(Q^2/m^2_c)$ terms are not resummed. Another short-coming is that the
NLO FFN coefficient functions are not available for the CC processes. 
Since the CC reactions at HERA are at high-$Q^2$ we use ZM-VFN coefficient 
functions. In practice the $\chi^2$ for these fits is not bad. 
The main difference between FFN and VFN fits is in sensitivity to the charm 
quark mass $m_c$, with the FFN fits preferring a low value $m_c=1.35~$GeV and 
the GM-VFN fits favouring a higher value $m_c=1.45~$GeV.

\begin{figure}[tbp]
\vspace{-0.7cm} 
\centerline{
\epsfig{figure=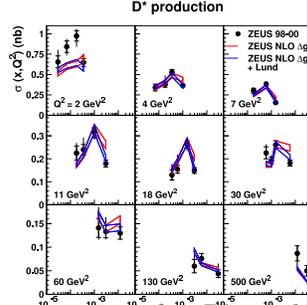,height=5.5cm}
}
\vspace{-2.0cm}
\caption {Double differential cross-sections for $D^*$ production. The red lines
 show the predictions of the ZEUS-S-13 NLO PDF fit using the Petersen 
fragmentation function for the $D^*$, whereas the blue lines show these 
predictions using the Lund fragmentation function.}
\label{fig:9xsecns}
\end{figure}

\begin{figure}[tbp]
\vspace{-2.0cm} 
\centerline{
\epsfig{figure=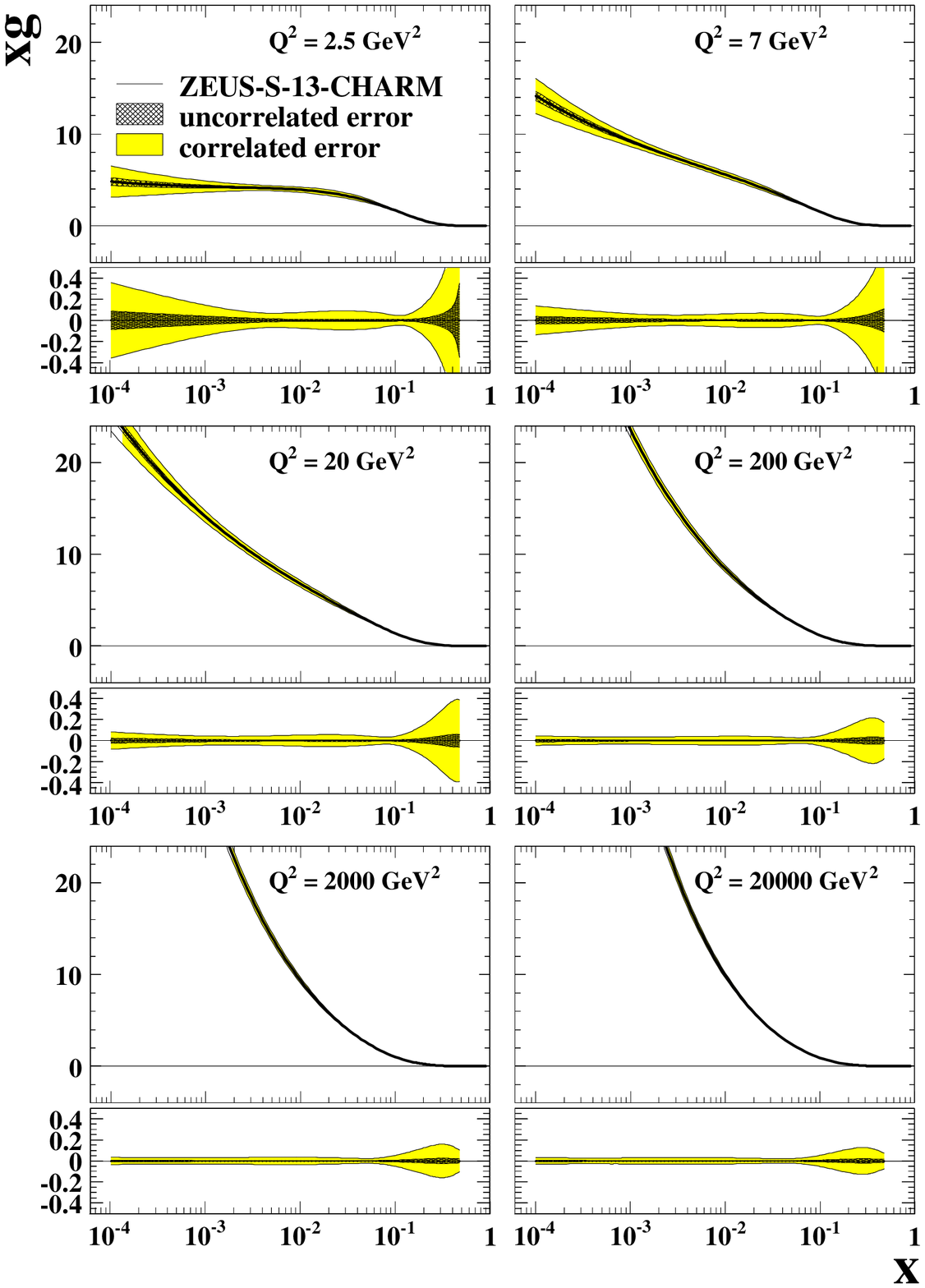 ,width=0.5\textwidth,height=5.5cm}
\epsfig{figure=mandy_coopersarkar1.fig7.eps ,width=0.5\textwidth,height=5.5cm}}
\vspace{-0.5cm}
\caption {The gluon PDF and its fractional uncertainties for various $Q^2$ bins
Left: before $D^*$ cross-section data are input to the ZEUS-S-13 fit. Right: after $D^*$ cross-section data are input to the ZEUS-S-13 fit}
\label{fig:gluonffnzs13}
\end{figure}

The small impact of the heavy flavour data on the global fit may be because 
we are not using the charm data optimally. 
$F_2^{c\bar{c}}$ is a quantity extracted from $D^*$ cross-sections by quite a 
large extrapolation. It would be better to fit to those cross-sections 
directly. The evaluation of the theoretical predictions involves running the 
NLO programme HVQDIS for each iteration of the fit. However, one can shorten 
this process by using the 
same method as was used for the ZEUS-JETS fit~\cite{zeusj}.  The
PDF independent subprocess cross-sections are output onto a grid, 
such that they can simply 
be multiplied by the PDFs at each iteration. 
The data used are the nine double differential cross-section measurements of
$d^2\sigma(D^*)/dQ^2dy$~\cite{dstar2003}, see Fig.~\ref{fig:9xsecns}.
When fitting $D^*$ cross-sections, 
as opposed to and inclusive quantity like $F_2^{c\bar{c}}$, one must use the 
FFN scheme since the prediction grids are calculated using HVQDIS. 
This means that we cannot use ZEUS high-$Q^2$ data. 
Hence we chose to use the ZEUS-S global 
fit~\cite{zeus-s} as the basis for our fit, with a cut-off $Q^2 < 3000$GeV$^2$.
The parametrisation was slightly modified to free the mid-$x$ 
gluon parameter $p_5(g)$ and the low-$x$ valence parameter $p_2(u)= p_2(d)$, 
such that the parametrization is like that of the ZEUS-JETS and ZEUS-pol fits.
This fit is called ZEUS-S-13.
Figure~\ref{fig:gluonffnzs13} 
shows the difference in the gluon PDF uncertainties, before and afer the 
$D^*$ cross-sections were input to the ZEUS-S-13 global fit. 
Disappointingly the uncertainty on the gluon is NOT much improved.

Should we have expected much improvement? 
There are two aspects of the fit which could be improved. The predictiond for 
the $D^*$ cross-sections have more uncertainties than just the PDF 
parametrization. A further  uncertainty is introduced in the choice of the 
$c \to D^*$ fragmentation  The Petersen fragmentation function was used for 
the fit predictions. However, looking back at Fig~\ref{fig:9xsecns} we can 
see that the Lund fragmentation function seems to describe the data better.
To best exploit the charm data in future we need to address such aspects 
of our model uncertainty. Secondly, this study on the D* cross-sections used 
only the HERA-I charm data. We look forward to the 5-fold increase in 
statistics expected from HERA-II charm and beauty data.


\begin{footnotesize}




\begin{thebibliography}{99}
\bibitem{url} Slides: \\ 
\verb$http://indico.cern.ch/contributionDisplay.py?contribId=246&sessionId=30&confId=24657$
\bibitem{mrst}
A.D.~Martin et al.,
\newblock Eur. Phys.J{} {\bf C23},~73~(2002)\relax
\relax
\bibitem{mrst06}
A.D.~Martin et al.,
\newblock arxiv{} {\bf 0706},~0459~(2007)\relax
\relax
\bibitem{cteq}
J.~Pumplin et al.,
\newblock JHEP{} {\bf 0207},~012~(2002)\relax
\relax
\bibitem{zeus-s}
ZEUS Coll., S.~Chekanov et al.,
\newblock Phys. Rev{} {\bf D~67},~012007~(2003)\relax
\relax
\bibitem{cteq65}
Wu Ki Tung et al.,
\newblock JHEP{} {\bf 0702},~053~(2007)\relax
\relax
\bibitem{hqnew}
R.S.~Thorne,
\newblock Phys.Rev{} {\bf D73},~054019~(2006)\relax
\relax
\bibitem{hq}
R.S.~Thorne and R.G.~Roberts,
\newblock Phys.Rev{} {\bf D57},~6871~(1998)\relax
\relax
\bibitem{dstar2003}
ZEUS Coll., S.~Chekanov et al.,
\newblock Phys. Rev{} {\bf D~69},~012004~(2004)\relax
\relax
\bibitem{hvqdis3}
B.W.~Harris and J.~Smith,
\newblock Phys.Lett.{} {\bf B~359},~423~(1995)\relax
\relax
\bibitem{zeusj}
ZEUS Coll., S.~Chekanov et al.,
\newblock Eur.Phys.J{} {\bf C~42},~1~(2005)\relax
\relax
\bibitem{match}
W.J.~Marciano,
\newblock Phys.Rev.{} {\bf D~29},~5801~(1984)\relax
\relax
\bibitem{kunihiro}
K.~Nagano,
\newblock Proceedings of DIS2008\relax
\relax
\bibitem{hera2charm}
ZEUS Collaboration,  
\newblock ZEUS-prel-07-007/008\relax
\relax
\bibitem{hera2beauty}
ZEUS Collaboration, 
\newblock ZEUS-Prel-07-009\relax


\end{thebibliography}
%

\end{footnotesize}


\end{document}